\documentclass[onecolumn,showpacs,epsfig,prd,preprintnumbers,amsmath,amssymb]{revtex4}
\usepackage{amsmath}
\usepackage{graphicx}
\usepackage{epsfig}
\usepackage{epsf}
\usepackage{graphicx}
\usepackage{graphics}
\usepackage{dcolumn}
\usepackage{bm}
\def \jp {J/\psi}
\def \llb {\Lambda\bar\Lambda}
\begin{document}
\title{\boldmath Measurement of the asymmetry parameter for the decay $\bar\Lambda\to\bar p\pi^+$}
\author{M.~Ablikim$^{1}$,              J.~Z.~Bai$^{1}$,
Y.~Bai$^{1}$, Y.~Ban$^{11}$, X.~Cai$^{1}$, H.~F.~Chen$^{16}$,
H.~S.~Chen$^{1}$,              H.~X.~Chen$^{1}$, J.~C.~Chen$^{1}$,
Jin~Chen$^{1}$
% ...
\\
\vspace{0.2cm} (BES Collaboration)\\
{\it
$^{1}$ Institute of High Energy Physics, Beijing 100049, People's Republic of China\\
% ...
$^{11}$ Peking University, Beijing 100871, People's Republic of China\\
% ...
$^{16}$ University of Science and Technology of China, Hefei 230026, People's Republic of China\\
% ...
}
\date{} }

\begin{abstract}
  Based on a sample of $58\times10^6J/\psi$ decays collected with the
  BESII detector at the BEPC, the $\bar\Lambda$ decay parameter
  $\alpha_{\bar\Lambda}$ for $\bar\Lambda\to \bar p \pi^+$ is measured
  using about 9000 $J/\psi\to\Lambda\bar\Lambda\to p \bar p \pi^+\pi^-$
  decays. A fit to the joint angular distributions yields
  $\alpha_{\bar\Lambda}(\bar\Lambda\to \bar
  p\pi^+)=-0.755\pm0.083\pm0.063$, where the first error is
  statistical, and the second systematic.
\end{abstract}
 \maketitle
\section{Introduction}
Nonleptonic hyperon decays were very important in early studies of
parity violation in particle physics~\cite{Lee&yang}. In a hyperon
decay, $Y\to B\pi$ ($Y$:hyperon, $B$:baryon), the angular
distribution of the baryon in the center-of-mass (CM) system of
$Y$ takes the form
$\textstyle\frac{\textit{dN}}{\textit{$d\Omega$}}\propto 1+
\textit{$\alpha_Y\vec{P}_Y\cdot\hat{p}_d$}$, where
$\textit{$\vec{P}_Y$}$ is the polarization vector of the hyperon,
$\textit{$\hat{p}_d$}$ is the momentum unit vector of the baryon
and $\alpha_Y$ is the hyperon decay parameter, which characterizes
the parity violation in hyperon nonleptonic decays. For the decay
$\Lambda\to p\pi^-$, the decay parameter is
$\alpha_\Lambda=0.642\pm 0.013$~\cite{pdg08}; for the counterpart
$\bar\Lambda\to\bar p\pi^+$, the decay parameter was only measured
by the DM$2$ Collaboration with large uncertainty:
$\alpha_{\bar\Lambda}=-0.63\pm0.13$ \cite{dm2}.

A precise measurement of the $\bar\Lambda$ decay parameter allows a
more sensitive
search for $CP$ symmetry violation in $\Lambda$ decays
\cite{E756,pais}. A $CP$-odd observable, $A$, may be defined as
\begin{equation}\label{cpa}
A={\alpha_{\Lambda}+\alpha_{\bar \Lambda}\over
\alpha_\Lambda-\alpha_{\bar \Lambda}}.
\end{equation}
If $CP$ is conserved, $\alpha_\Lambda=-\alpha_{\bar \Lambda}$ and
this observable vanishes; while a nonzero value of $A$ implies
evidence for $CP$ asymmetry in $\Lambda$ decays.  Previous
analyses to search for $CP$ asymmetry in $\Lambda$ nonleptonic
decays have been performed at $p\bar p$ colliders by the R608
\cite{r608} and PS185 \cite{ps185} Collaborations, and at an
$e^+e^-$ collider by the DM2 Collaboration \cite{dm2}, but the
precisions of the measurements are limited by low statistics.

A precise $\bar\Lambda$ decay parameter is also essential in
determining the $\bar\Omega^+$ or $\bar\Xi^+$ decay parameters.
Nonpolarized $\bar\Omega^+$ or $\bar\Xi^+$ decays can produce
polarized $\bar\Lambda$ particles, so in the $\bar\Lambda$ rest
frame, the angular distribution of the final state antiproton,
takes the form ${d N\over d\cos\theta}\propto 1+\alpha_{\bar
\Omega}\alpha_{\bar\Lambda}\cos\theta$ \cite{omega}. To extract
$\alpha_{\bar \Omega}$ from the product $\alpha_{\bar
\Omega}\alpha_{\bar\Lambda}$ the value of $\alpha_{\bar\Lambda}$
is required. A similar argument holds for $\bar\Xi^+$ decays.

Pair production of $\Lambda \bar\Lambda$ in $J/\psi$ decays allows
one to study the $\bar\Lambda$ decay parameter. Although the
$\Lambda$ and $\bar\Lambda$ particles are nonpolarized in $\jp$
decays, their helicities are correlated by helicity conservation.
Hence, the $\bar\Lambda$ decay parameter can be extracted from the
helicity correlation between $\Lambda$ and $\bar\Lambda$.
Experimentally, this decay is very clean and can be reconstructed
with high efficiency by selecting events with four charged tracks.
The $58\times10^6$ $J/\psi$ decays used for this analysis were
taken with the BESII detector at the BEPC storage ring at a
center-of-mass energy corresponding to \textit{$M_{J/\psi}$}.
They offer an opportunity to measure a more precise $\bar\Lambda$
decay parameter.

\section{The BES experiment}

The Beijing Spectrometer (BES) detector is a conventional
solenoidal magnet detector that is described in detail in Ref.
\cite{detector}; BESII is the upgraded version of the BES detector
\cite{detector1}. A $12$-layer vertex chamber (VTC) surrounding
the beam pipe provides trigger and track information. A 40-layer
main drift chamber (MDC), located radially outside the VTC,
provides trajectory and energy loss ($dE/dx$) information for
charged tracks over $85$\% of the total solid angle. The momentum
resolution is $\sigma_p/p=0.017\sqrt{1+p^2}$ ($p$ ~in GeV/$c$),
and the $dE/dx$ resolution for hadron tracks is $\sim8$ \%. An
array of $48$ scintillation counters surrounding the MDC measures
the time-of-flight (TOF) of charged tracks with a resolution of
$\sim200$ ps for hadrons. Radially outside the TOF system is a
$12$ r.l., lead-gas barrel shower counter (BSC). This measures the
energies of electrons and photons over $\sim80$\% of the total
solid angle with an energy resolution of
$\sigma_E/E=22\%/\sqrt{E}$ ($E$ in GeV). Outside of the solenoidal
coil, which provides a $0.4$ T magnetic field over the tracking
volume, is an iron flux return that is instrumented with three
double layers of counters that identify muons of momentum greater
than $0.5$ GeV/$c$.

A GEANT$3$ based Monte Carlo (MC) program with detailed
consideration of detector performance (such as dead electronic
channels) is used to simulate the BESII detector. The consistency
between data and Monte Carlo has been carefully checked in many
high purity physics channels, and the agreement is quite
reasonable~\cite{compare}.
Several $J/\psi\to\Lambda\bar\Lambda$ MC samples are generated and
used for determining the detection efficiency, determining the
normalization factor in background subtraction, and performing an
input-output check.

\section{Event selection}
\begin{figure}
\begin{center}

  \parbox{0.99\textwidth}{\includegraphics[width=8cm,height=8cm]{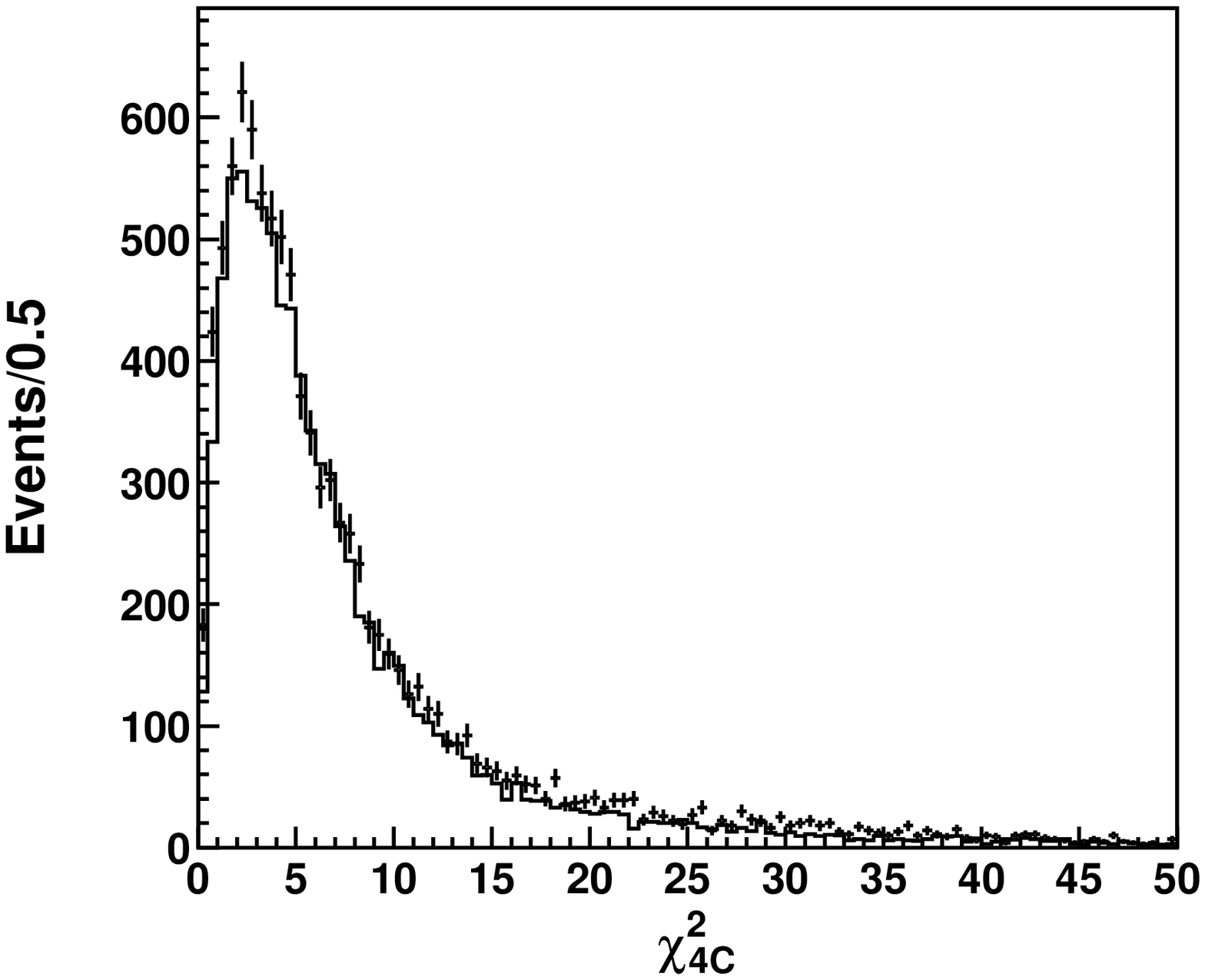}\includegraphics[width=8cm,height=8cm]{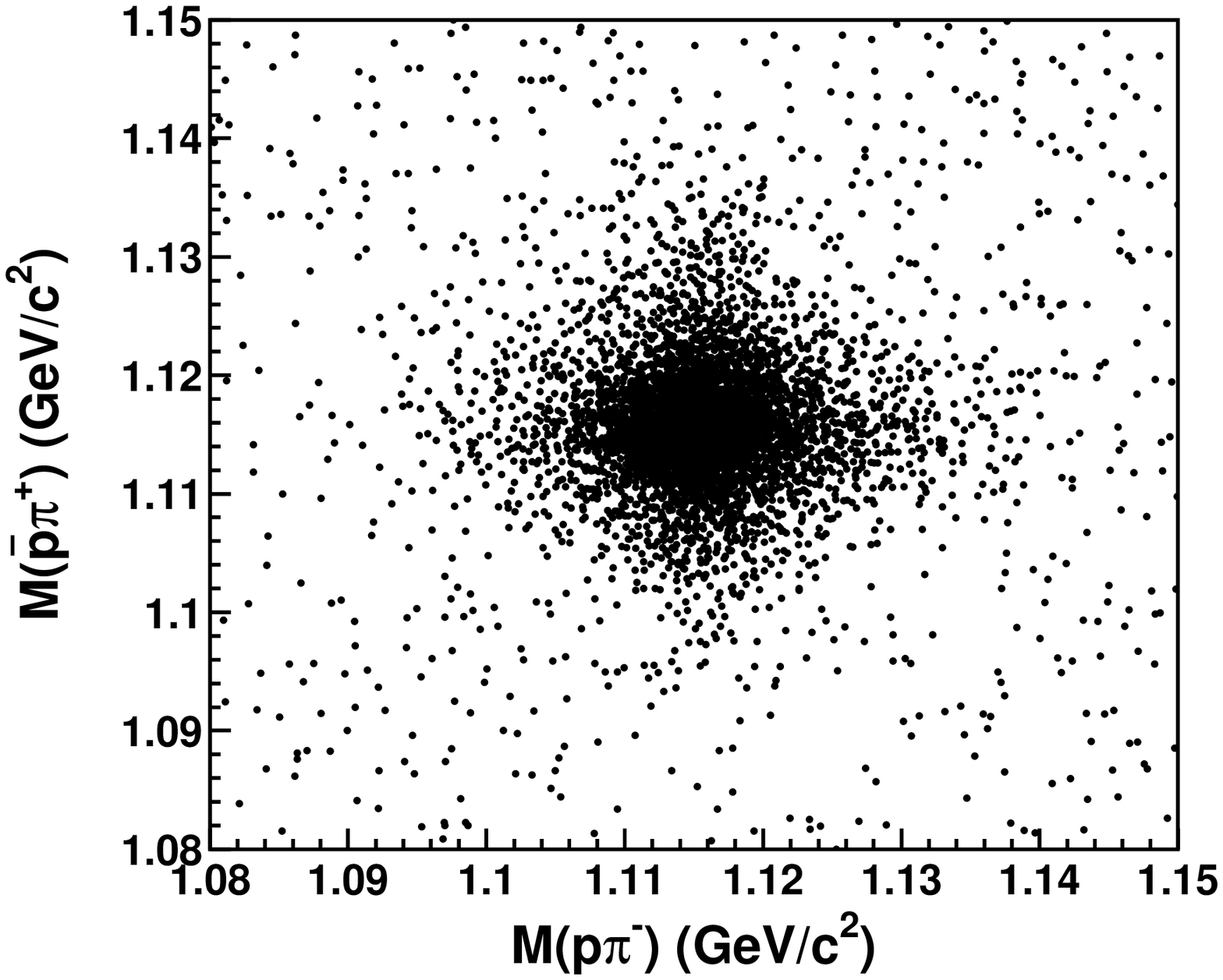}}
  \parbox{1.0\textwidth}{\vspace{-5cm}\hspace{-1cm}(a)\hspace{6cm}(b)}
  \parbox{0.6\textwidth}{\caption{Distributions for selected
      $J/\psi\to\Lambda\bar\Lambda$ candidates. (a) $\chi^2_{4C}$.
      Dots with error bars are data; histogram is for MC simulation.
      (b) Scatter plot of $m_{p\pi^-}$ versus $m_{\bar
        p\pi^+}$.\label{scaterplot}}}
\end{center}
\end{figure}

\begin{figure}[htbp]
\begin{center}
\vspace*{0.5cm}
\parbox{1.0\textwidth}{\includegraphics[width=8cm,height=8cm]{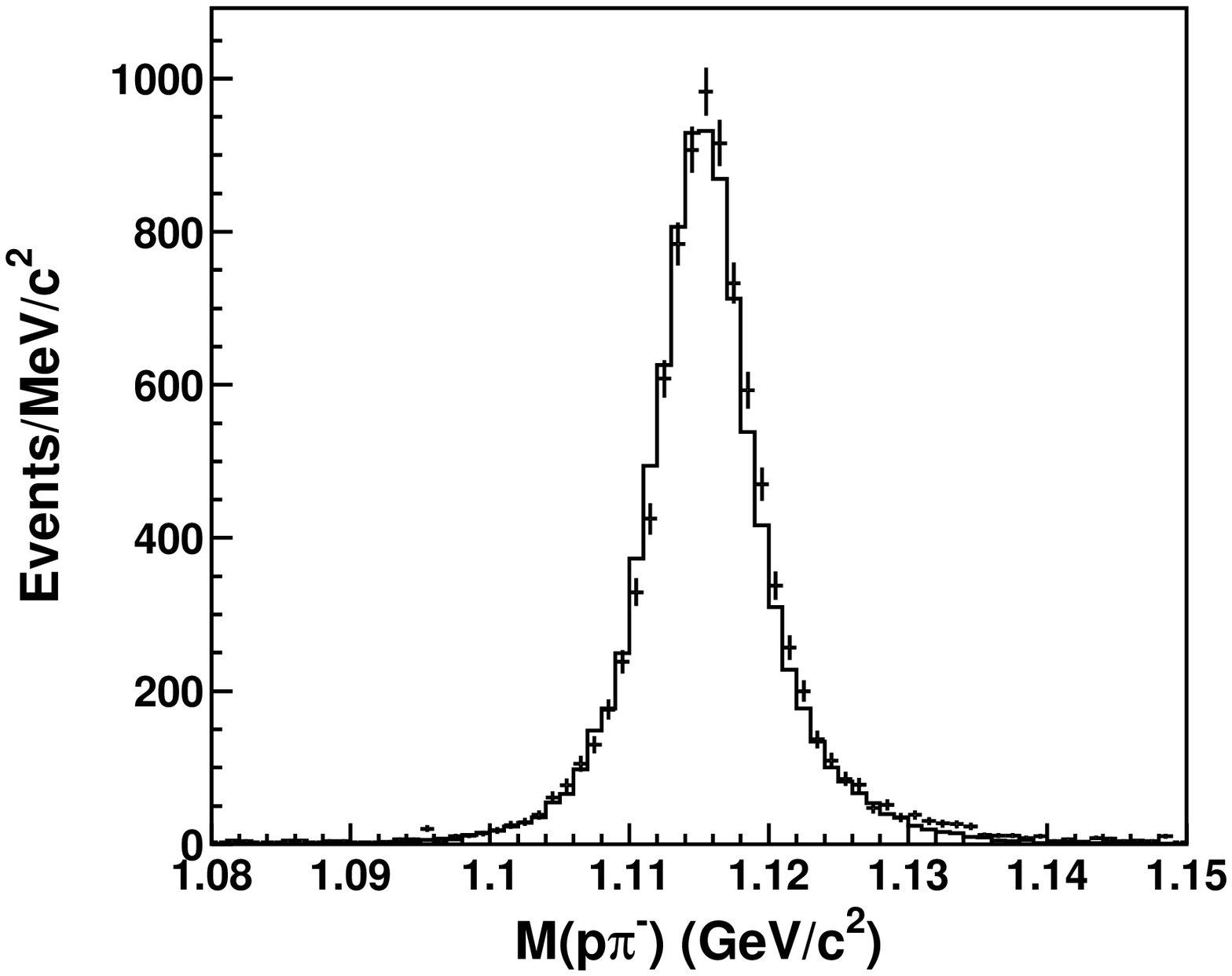}
\hspace{0cm}\includegraphics[width=8cm,height=8cm]{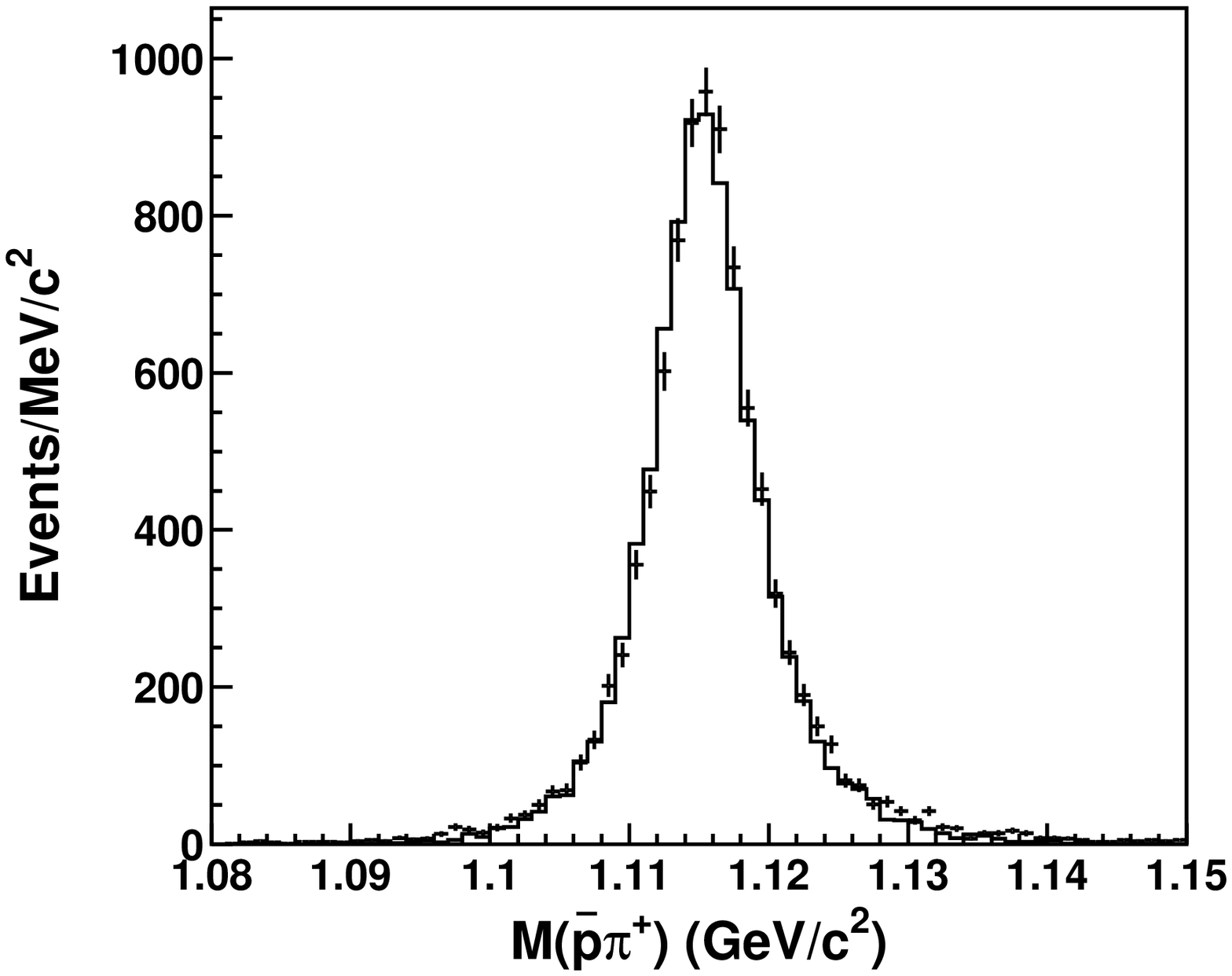}}
\parbox{1.0\textwidth}{\vspace{-5cm}\hspace{1cm}(a)\hspace{5cm}(b)}
\parbox{0.8\textwidth} { \caption{Comparisons of invariant mass
    distributions between data and MC. Histograms are
    normalized MC; points with error bars are data. (a) $m_{p\pi^-}$ and
    (b) $m_{\bar p\pi^+}$. \label{mlambda}}}
\end{center}
\end{figure}

The $\Lambda$ is reconstructed under the assumption of $\Lambda
\to p\pi^-$ decay, and $\bar\Lambda$ under the $\bar\Lambda \to
\bar p\pi^+$ hypothesis. A candidate track is required to have a
good helix fit, the polar angle must satisfy
$|\cos\theta_{\textnormal{ch}}|<0.8$, and transverse momentum
$p_{xy}>0.07~$GeV/c. Events are required to have four charged
tracks with total charge of zero. Protons and antiprotons are
identified using TOF and $dE/dx$ information with the requirement
that their confidence levels be larger than $0.01$. A four
constraint (4C) kinematic fit is applied under the $p\bar
p\pi^+\pi^-$ hypothesis, and $\chi^2_{4C}<20$ is required. A
comparison of $\chi^2_{4C}$ distributions of data and MC
simulation is shown in Fig. \ref{scaterplot} (a). The scatter plot
of $m_{\bar p\pi^+}$ versus $m_{p\pi^-}$ is shown in Fig.
\ref{scaterplot} (b), and $\jp\to\llb$ is clearly seen. The
$\Lambda$ and $\bar\Lambda$ are selected by requiring
$|M_{p\pi^-}-1.1156|<0.015 ~\textrm{GeV/c}^2$ and $|M_{\bar
  p\pi^+}-1.1156|<0.015 ~\textrm{GeV/c}^{2}$.  The mass distributions
of $p\pi^-$ and $\bar p\pi^+$ are shown in Fig. \ref{mlambda}.

After applying the above selection criteria,  $8997$
$J/\psi\to\Lambda\bar\Lambda$ events are selected. The angular
distribution of $\Lambda$ in the $J/\psi$ rest frame is found to be
consistent with the dedicated analysis in Ref. \cite{bes21} within
uncommon statistical errors.

Backgrounds are studied with MC simulations. The main backgrounds come
from $J/\psi \rightarrow \Sigma^0\bar{\Sigma}^0 \rightarrow
2\gamma\Lambda\bar{\Lambda} \rightarrow 2\gamma p\pi^-\bar{p}\pi^+$
(28 events), $J/\psi \rightarrow \Lambda\bar{\Sigma}^0 \rightarrow
\gamma\Lambda\bar{\Lambda} \rightarrow \gamma p\pi^-\bar{p}\pi^+$ (36
events), $J/\psi \rightarrow \bar{\Lambda}\Sigma^0 \rightarrow
\gamma\Lambda\bar{\Lambda} \rightarrow \gamma p\pi^-\bar{p}\pi^+$ (37
events), $J/\psi\to p\pi^-p\pi^+$ (10 events),
$J/\psi\to\Delta^{++}\Delta^{--}$ (5 events),
$J/\psi\to\Delta^{++}\bar p\pi^-$ (1 events), and
$J/\psi\to\Delta^{--}p\pi^+$ (8 events). The fraction of remaining
background events in the data sample is about $1.4\%$. Background
decays to intermediate states including $\Lambda$ or $\bar\Lambda$ are
generated with the helicity amplitude method \cite{zhongbin}, in which
hyperon decay parameters are set according to PDG values \cite{pdg08}.
The effect of the background contamination on the physics results is
included as one source of systematic error.

\section{Fit to data}
\begin{figure}
\begin{center}
\includegraphics[height=6cm]{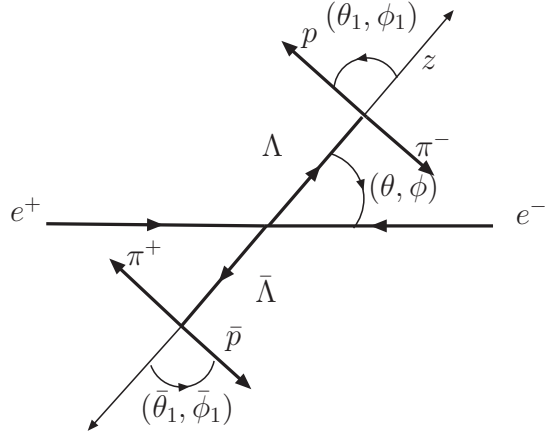}
\centering
\parbox{0.6\textwidth}{
\caption{Definition of the helicity frame for $J/\psi \rightarrow
\Lambda \bar{\Lambda} \rightarrow
 p\pi^-\bar{p}\pi^+$.\label{frame}}}
\end{center}
\end{figure}

The angles used in this analysis are defined in the
helicity frame, shown in Fig. \ref{frame}. For $J/\psi \rightarrow
\Lambda\bar{\Lambda}$, the $z$ axis of the $J/\psi$ rest frame is
along the $\Lambda$ out-going direction, and the solid angle
$\Omega_0(\theta,\phi)$ is between the $e^+$ direction and the
$\Lambda$ out-going direction. For $\Lambda\rightarrow p\pi^{-}$, the
solid angle of the daughter particle $\Omega_i(\theta_i,\phi_i)$ is
referred to the $\Lambda$ rest frame, and the $z$ axis is also the
$\Lambda$ out-going direction. For $\bar\Lambda$ we use symbols with
bars.

For $J/\psi \rightarrow
 \Lambda \bar{\Lambda} \rightarrow
 p\pi^-\bar{p}\pi^+$, the partial decay rate is \cite{zhongbin}
\begin{eqnarray}\label{ampllb}
\frac{d\sigma}{d\Omega} &\propto&
(1-\alpha)\sin^2\theta[1+\alpha_\Lambda\alpha_{\bar{\Lambda}}(\cos\theta_1\cos\bar{\theta}_1+\sin\theta_1\sin\bar{\theta}_1\cos(\phi_1+\bar{\phi}_1))]\nonumber\\
&-&(1+\alpha)(1+\cos^2\theta)(\alpha_\Lambda\alpha_{\bar{\Lambda}}\cos\theta_1\cos\bar{\theta}_1-1),
\end{eqnarray}
where $d\Omega=d\Omega_0d\Omega_1d\bar\Omega_1$, $\alpha$ is the
angular distribution parameter for $\Lambda$, and
$\alpha_{\Lambda}(\alpha_{\bar\Lambda})$ is the
$\Lambda~(\bar\Lambda)$ decay parameter. Equation. (\ref{ampllb})
only allows one to obtain the product
$\alpha_\Lambda\alpha_{\bar\Lambda}$ from a fit to data. To
extract the value of $\alpha_{\bar\Lambda}$, one needs to fix the
$\Lambda$ decay parameter $\alpha_{\Lambda}$, say, at the world
average value $\alpha_{\Lambda}=0.642$ \cite{pdg08}. To determine
the $CP$-odd observable defined in Eq. (\ref{cpa}), one may make a
replacement
\begin{equation}
\alpha_\Lambda\alpha_{\bar\Lambda}={A-1\over
A+1}\alpha_{\Lambda}^2
\end{equation}
in Eq. (\ref{ampllb}).

An unbinned maximum likelihood method is used to fit the data. As
widely used in partial wave analyses\cite{pwa}, a normalized
probability density function for $J/\psi\to\Lambda\bar\Lambda\to
p\pi^-\bar p \pi^+$ is defined as
\begin{equation}\label{probability}
\textrm{Prob}(\Omega,\alpha,\alpha_{\bar\Lambda})={d\sigma /
d\Omega \over \sigma},
\end{equation}
where $\Omega=(\Omega_0,\Omega_1,\bar\Omega_1)$, and $\alpha$ and
$\alpha_{\bar\Lambda}$ are parameters to be determined. $d\sigma /
d\Omega$ is given by Eq. (\ref{ampllb}). $\sigma$ is the total cross
section given by
\begin{equation}\label{crosssection}
\sigma=\int {d\sigma \over d\Omega} \epsilon(\Omega)d\Omega,
\end{equation}
where $\epsilon(\Omega)$ is the detection efficiency. The total
cross section $\sigma$ can be determined by MC numerical
integration with a phase space generator over the allowed
kinematic region of $J/\psi\to\Lambda\bar\Lambda\to p\pi^-\bar p
\pi^+$, i.e.
\begin{equation}\label{crossectionmc}
\sigma={1 \over N_{MC}}\sum_{i=1}^{N_{MC}} \left\{{d\sigma \over
d\Omega}\right\}_i,
\end{equation}
where $N_{MC}$ is the number of selected MC events.

The maximum likelihood function is given by the joint probability
density of the selected $J/\psi\to\Lambda\bar\Lambda\to p\pi^-
\bar p \pi^+$ events
\begin{equation}\label{likehood}
\mathcal{L}=\prod_{i=1}^{N_{evt}}\textrm{Prob}(\Omega,\alpha,\alpha_{\bar\Lambda}),
\end{equation}
where $N_{evt}$ is the number of selected data events. To
determine the unknown parameters $\alpha$ and
$\alpha_{\bar\Lambda}$, the function $S$, which is defined as
\begin{equation}\label{minimum}
\mathcal{S}=-\ln\mathcal{L},
\end{equation}
is minimized.

\begin{figure}
\begin{center}
\parbox{1\textwidth}{\includegraphics[width=8cm,height=8cm]{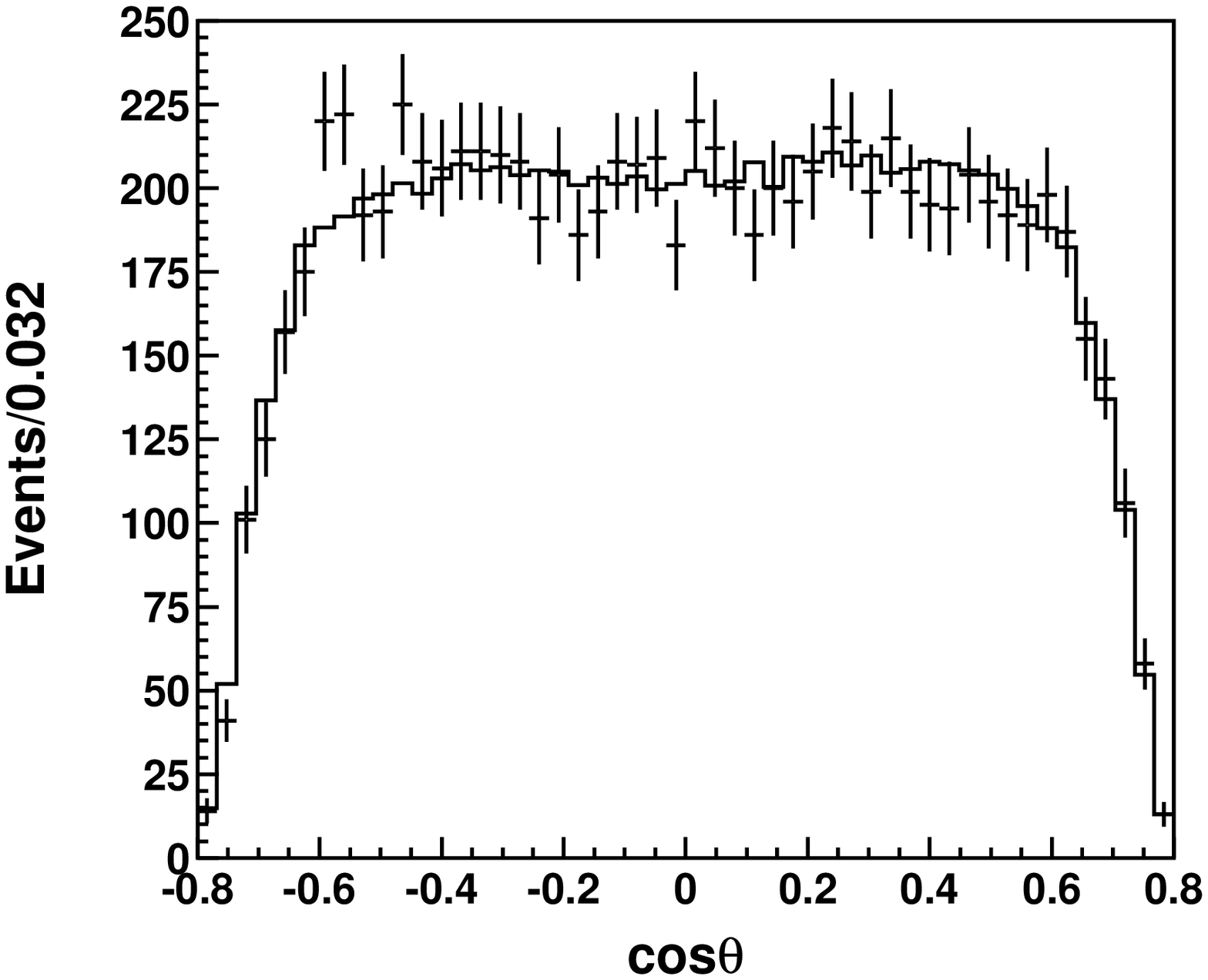}\includegraphics[width=8cm,height=8cm]{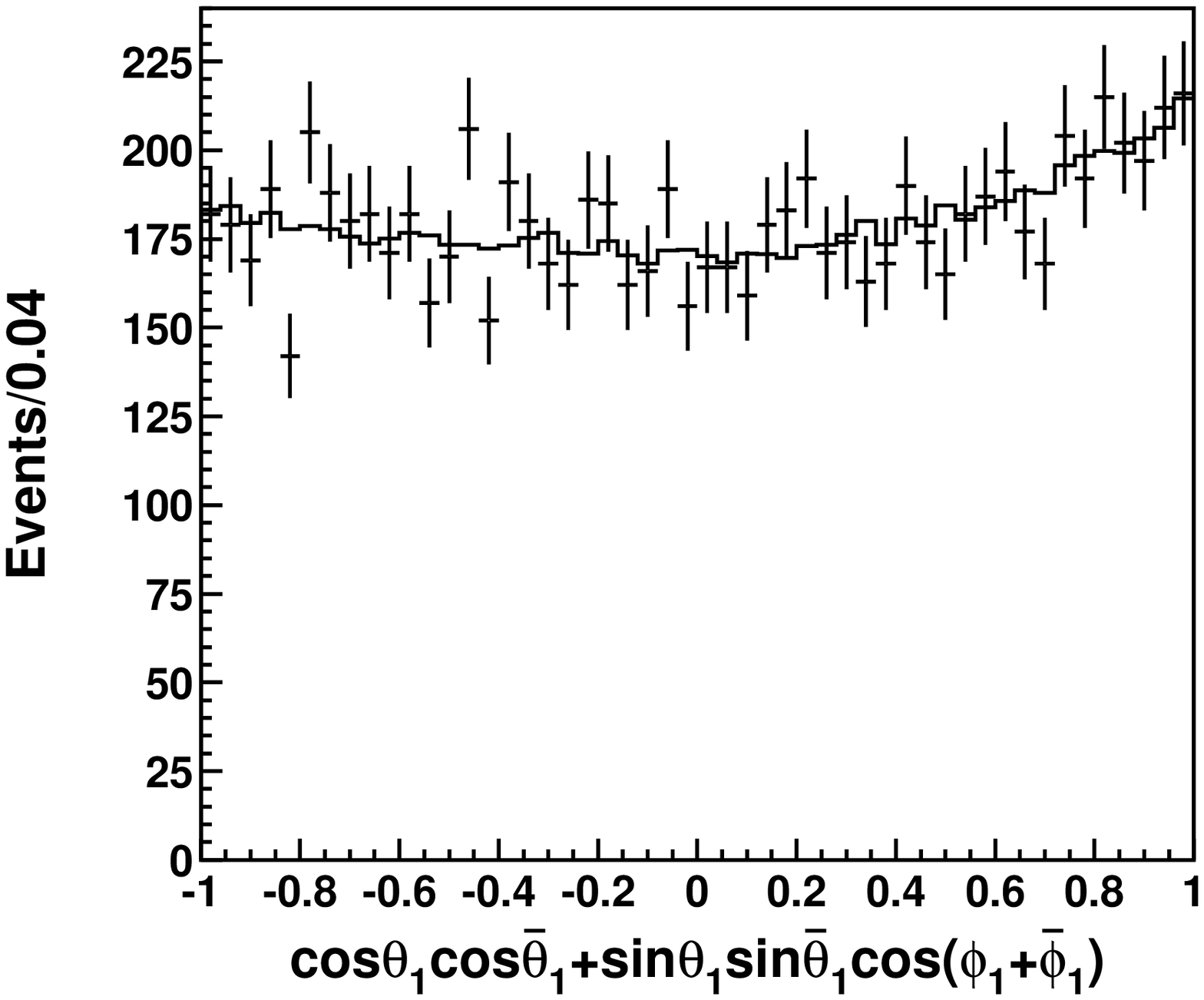}}
\parbox{1.0\textwidth}{\vspace{-5cm}\hspace{1cm}(a)\hspace{5cm}(b)}
\parbox{0.9\textwidth}{
\caption{Comparison between data (dots with error bars) and fit results
  (histograms). (a) Distribution of $\cos\theta$ of $\Lambda$ in
  $J/\psi$ rest frame. (b) Distribution of
    $\cos\theta_1\cos\bar\theta_1+\sin\theta_1\sin\bar\theta_1\cos(\phi_1+\bar\phi_1)$.
\label{asydis}}}
\end{center}
\end{figure}

To check the goodness of fit in our analysis, we define
\begin{equation}\label{chisqcheck}
\chi^2=\sum_{i=1}^N \frac{(n_i^{DT}-n_i^{MC})^2}{n_i^{DT}},
\end{equation}
over the scatter plot of $\cos\theta$ versus
$|\cos\theta_1\cos\bar\theta_1+\sin\theta_1\sin\bar\theta_1\cos(\phi_1+\bar\phi_1)|$.
Here $N$ is the number of cells, $n_i^{DT}$ and $n_i^{MC}$ are the
numbers of events in the $i$th cell of the scatter plot for data
and MC simulation, respectively. Such a variable should be
distributed according to the $\chi^2$ distribution with $ndf=N-K$
degrees of freedom, where $K=2$ is the number of parameters to be
determined in our fit. In this experiment, data and MC events are
divided into $8\times 10$ cells in the scatter plots.

Fitting the data using Eq. (\ref{minimum}), we obtain
\begin{equation}
\alpha = 0.70 \pm 0.06,
\end{equation}
\begin{equation}
\alpha_{\bar\Lambda} =-0.755\pm0.083,
\end{equation}
while fitting the data using Eq. (\ref{minimum}) with parameters
$\alpha$ and $A$, we obtain
\begin{equation}
A =-0.081 \pm 0.055,
\end{equation}
where the errors are statistical only. Comparisons between the data and
the fit results are shown in Fig. \ref{asydis} (a) and (b). The fit
yields $\chi^2/ndf=78.59/(80-2)=1.01$.
If we force $A=0$, we
obtain $\alpha_{\Lambda}=-\alpha_{\bar\Lambda}= 0.696 \pm 0.038$,
where the error is statistical only.
\section{INPUT-OUTPUT CHECK}

To validate the fitting procedure, a MC sample of $2\times10^6$
$\jp\to\Lambda\bar\Lambda\to p\bar p \pi^+\pi^-$ events is
produced according to Eq. (\ref{ampllb}). The input parameters are
$\alpha=0.62$ and $\alpha_\Lambda=-\alpha_{\bar\Lambda}=0.642$.
The MC sample is required to pass the same selection criteria as
used for data selection, and the same fitting procedure is applied
to the selected events with $\alpha_\Lambda$ fixed to $0.642$. The
fit yields $\alpha=0.612\pm0.010$ and
$\alpha_{\bar\Lambda}=-0.640\pm0.013$,  consistent with the input
values to within the $1\sigma$ statistical errors.

\section{SYSTEMATIC ERRORS}
\subsection{Background contamination}
Contamination from background channels is studied using MC
samples, including
$\jp\to\Sigma^0\bar\Sigma^0,~\Lambda\bar\Sigma^0+c.c.,~\Delta^{++}\Delta^{--},~\Delta^{++}\bar
p\pi^-,~\Delta^{--} p\pi^+$ and $\bar pp\pi^+\pi^-$. Background
channels with $\Lambda/\bar\Lambda$ intermediate states are
generated with full helicity amplitude information, and the decay
parameters for $\Lambda,~\bar\Sigma^0$ and their antiparticles are
fixed at the world average values \cite{pdg08}.  These samples are
required to pass the same selection criteria as used for the the
signal channel $J/\psi\to\Lambda\bar\Lambda\to\bar p p\pi^+\pi^-$.
The selected background events are subtracted from the data in the
fit. The difference of the $\bar\Lambda$ decay parameter
$\alpha_{\bar\Lambda}$ from the fit without background events,
$\delta\alpha_{\bar\Lambda}= 0.021$, is taken as the uncertainty
from background contamination.

\subsection{\boldmath Decay parameter $\alpha_\Lambda$ }
The systematic error due to the uncertainty on the $\Lambda$ decay
parameter~\cite{pdg08} is determined. In fitting to data, the
central value of $\alpha_{\bar\Lambda}$ is obtained by fixing
$\alpha_{\Lambda}=0.642$; the difference from the central value,
$\delta\alpha_{\bar\Lambda}=0.015$, is obtained by changing
$\alpha_\Lambda$ by 1 standard deviation ($\pm0.013$).

\subsection{MC simulation and detector response}
\begin{figure}
\begin{center}
\parbox{1\textwidth}{\includegraphics[width=8cm,height=8cm]{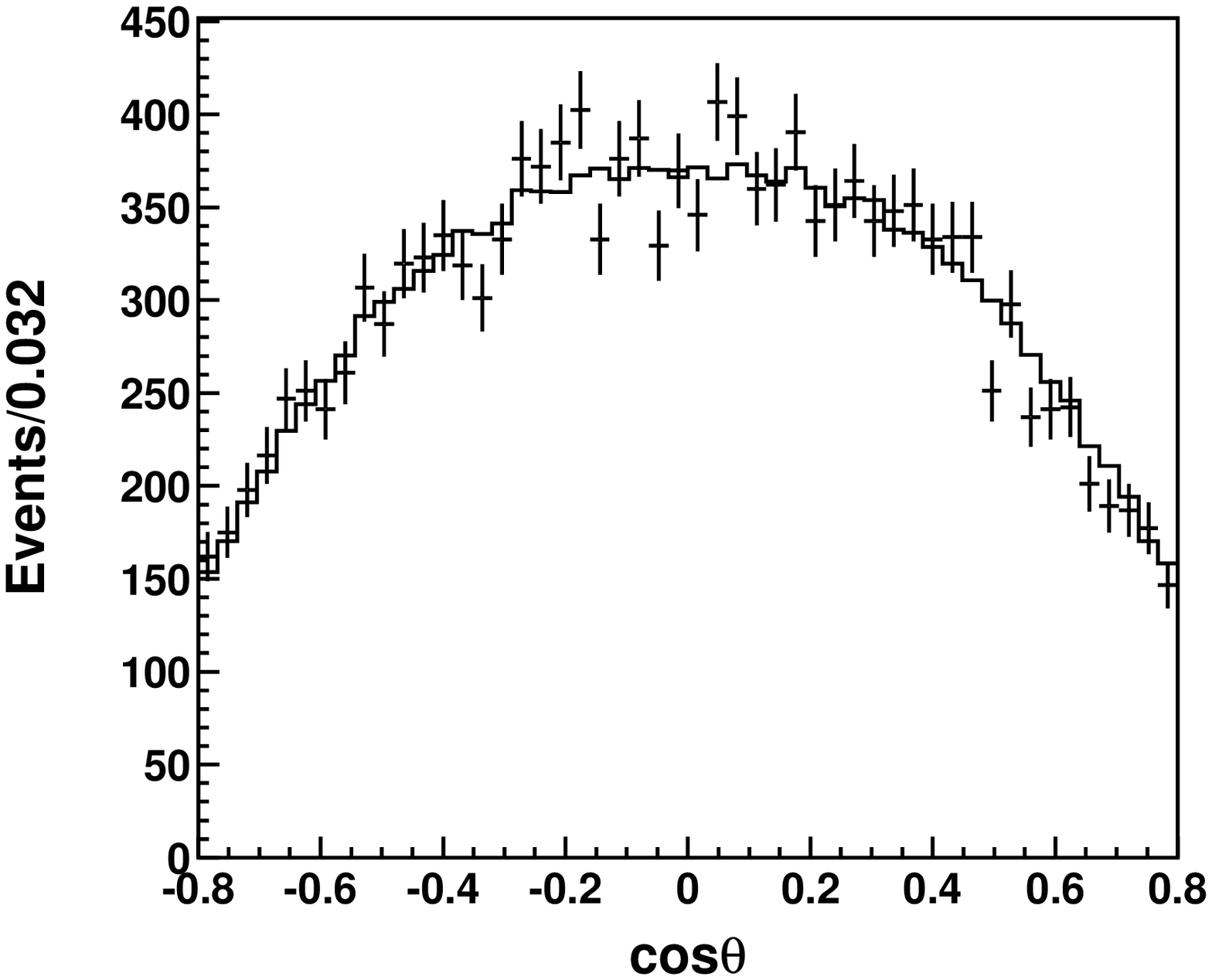}\includegraphics[width=8cm,height=8cm]{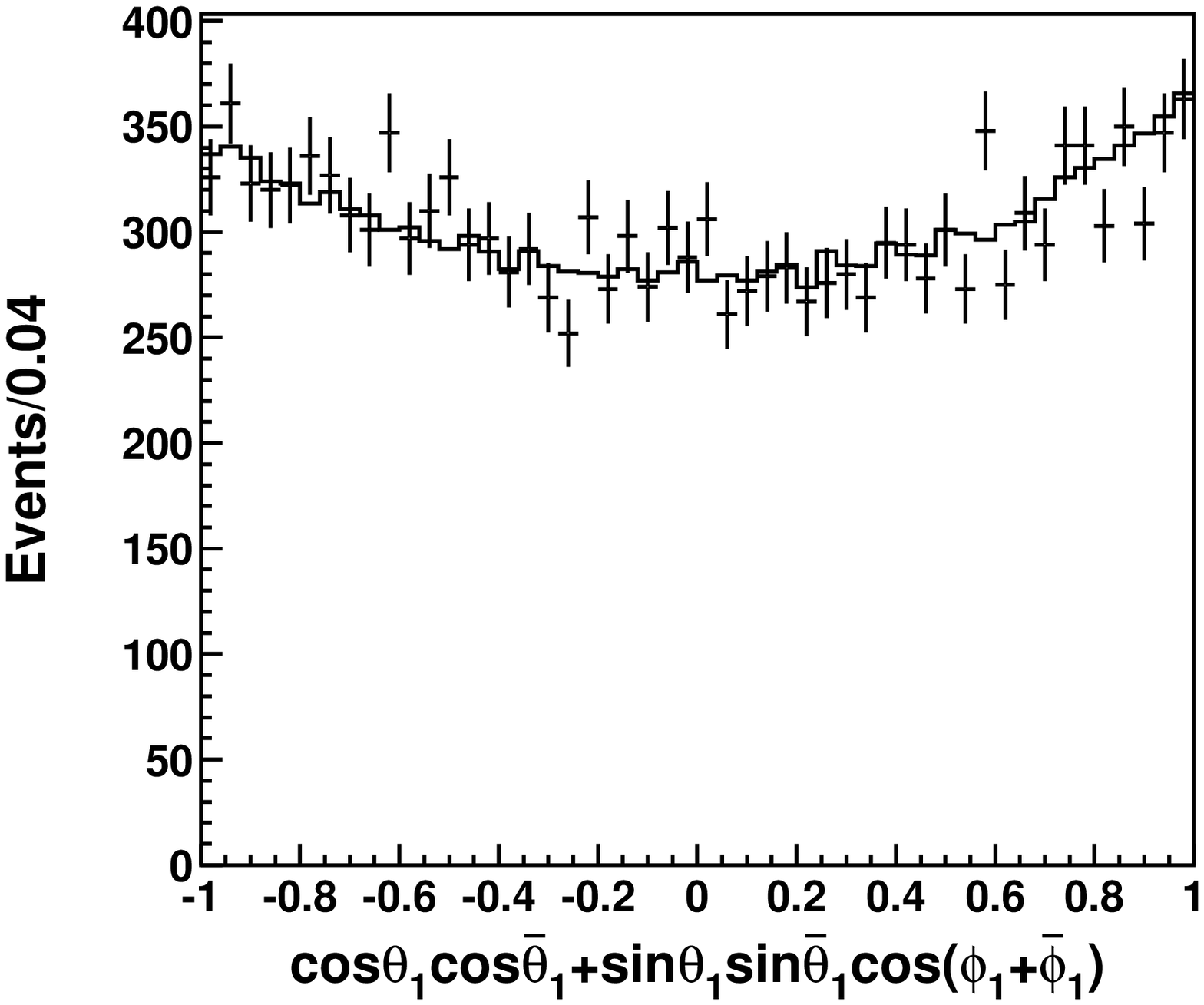}}
\parbox{1.0\textwidth}{\vspace{-5cm}\hspace{1cm}(a)\hspace{5cm}(b)}
\parbox{0.9\textwidth}{
\caption{Fitting results of the control sample $\jp\to~p\bar
  p\pi^+\pi^-$ used for determining the uncertainty associated with
  the MC simulation and detector response. Points with error bars are
  data, and the histograms are the fit results. (a) Distribution of $\cos \theta$. (b) Distribution of
  $\cos\theta_1\cos\bar\theta_1+\sin\theta_1\sin\bar\theta_1\cos(\phi_1+\bar\phi_1)$.\label{controlsample}}}
\end{center}
\end{figure}

The consistency between data and the MC simulation of the detector
response for $\jp\to\Lambda\bar\Lambda$ events can be determined using
the channel with the same final states $\jp\to p\bar p \pi^+\pi^-$
after rejecting $\Lambda$, $\bar\Lambda$, $\Delta^{++}$ and
$\Delta^{--}$ intermediate states.
Therefore fitting to this sample with Eq. (\ref{ampllb}) should
yield $\alpha_\Lambda\alpha_{\bar\Lambda}=0$. The difference from
zero gives the systematic error due to the MC simulation of the
detector response. In fitting this sample with Eq. (\ref{ampllb}),
angles for $\Lambda$ or $\bar\Lambda$ are replaced with ones for
the quasi-two body systems ($p\pi^-$ and $\bar p\pi^+$). The fit
yields $\alpha_\Lambda\alpha_{\bar\Lambda}=0.008\pm0.036$; the
uncertainty from MC simulation and detector response is taken as
$0.044$, the linear sum of the deviation and its uncertainty. The
resulting distributions of $\cos\theta$ and
$\cos\theta_1\cos\bar\theta_1+\sin\theta_1\sin\bar\theta_1\cos(\phi_1+\bar\phi_1)$
are shown in Fig. \ref{controlsample}.

\subsection{Hadron interaction model and wire resolution}
The systematic error associated with MC simulation of hadronic
interactions of final state particles with the detector is estimated
with two models, GCALOR \cite{gcalor} and FLUKA \cite{fluka}. The
central value is obtained with GCALOR model, while the difference
from the central value using FLUKA model,
$\delta\alpha_{\bar\Lambda}=0.005$, is taken as the systematic
error. The uncertainty associated with the simulation of the MDC
wire resolution is estimated with two different versions of MC
simulation software, which yields $\delta\alpha_{\bar\Lambda}=0.037$.

Table \ref{sys} summarizes all systematic errors. The total
systematic error is estimated to be
$\delta\alpha_{\bar\Lambda}=0.063$, assuming all the sources are
independent and summing them in quadrature.

\begin{table}
\begin{center}
\caption{Systematic errors in $\alpha_{\bar\Lambda}$.}\label{sys}
\begin{tabular}{lc}
\hline Backgrounds & $0.021$\\
 $\alpha_\Lambda$&$0.015$\\
MC simulation and detector response& 0.044\\
 Hadron model& 0.005 \\
 Wire resolution & 0.037 \\\hline
Total & $0.063$\\
\hline
\end{tabular}
\end{center}
\end{table}

\section{RESULT AND DISCUSSION}
Including systematic errors, we obtain
\begin{eqnarray}
\alpha_{\bar\Lambda}=-0.755\pm0.083\pm0.063,\nonumber\\
A=-0.081\pm0.055\pm0.059,\nonumber
\end{eqnarray}
where the first errors are statistical and the second systematic. The
comparison between DM2 \cite{dm2} and our results for $\alpha$ and $A$
is given in Table \ref{Alam}. They agree with each other within
errors, while BES has improved precision.

Our measurement gives the value of $CP$-odd variable
$A=-0.081\pm0.081$. The precision of this measurement is
insufficient to observe $CP$ violation at the level predicted by
the standard model: $A=-2.10\times10^{-5}$ in the
Kobayashi-Maskawa model or $A=-1.10\times10^{-4}$ in the Weinberg
model \cite{dm2}. The precision of this measurement is expected to
be highly improved at BESIII in the near future with $10^{10}$
$\jp$ decays accumulated.

\begin{table}
\begin{center}
\caption{Comparison of $\alpha_{\bar\Lambda}(\bar\Lambda\to\bar p\pi^+)$ and $A$ with DM2 measurements~\cite{dm2}.\label{Alam}}
\begin{tabular}{lcc}
\hline
  & $\alpha_{\bar\Lambda}(\bar\Lambda\to\bar p\pi^+)$ & $A$\\
\hline DM$2$ & $-0.63\pm0.13$ & $0.01\pm0.10$\\
This work & $-0.755\pm0.083\pm0.063$ & $-0.081\pm0.055\pm0.059$\\
\hline
\end{tabular}
\end{center}
\end{table}

\section{ACKNOWLEDGMENTS}
The BES Collaboration thanks the staff of BEPC and computing
center for their hard efforts. This work is supported in part by
the National Natural Science Foundation of China under Contracts
No. 10491300, No. 10225524, No. 10225525, No. 10425523, No.
10625524, No. 10521003, No. 10821063, No. 10825524 and No.
10225522 (Tsinghua University), the Chinese Academy of Sciences
under Contract No. KJ 95T-03, the 100 Talents Program of CAS under
Contracts No. U-11, No. U-24, No. U-25, and the Knowledge
Innovation Project of CAS under Contracts No. U-602, No. U-34
(IHEP), and the Department of Energy under Contract No.
DE-FG02-04ER41291 (U. Hawaii).

\end{document}